\documentclass[a4paper,UKenglish,cleveref, autoref]{lipics-v2019}
\usepackage{complexity}
\usepackage{upgreek}
\usepackage[ruled,vlined,titlenotnumbered]{algorithm2e}

\usepackage{tcolorbox}

\newcommand{\bd}{\mathrm{bd}}
\newtheorem{conjecture}[theorem]{Conjecture}

\title{Faster and simpler algorithms for finding large patterns in permutations} %TODO Please add

\titlerunning{Faster and simpler algorithms for finding large patterns in permutations}%optional, please use if title is longer than one line

\author{L\'aszl\'o Kozma}{Institut f\"ur Informatik, Freie Universit\"at Berlin}{laszlo.kozma@fu-berlin.de}{}{}%TODO mandatory, please use full name; only 1 author per \author macro; first two parameters are mandatory, other parameters can be empty. Please provide at least the name of the affiliation and the country. The full address is optional

\authorrunning{L.~Kozma}%TODO mandatory. First: Use abbreviated first/middle names. Second (only in severe cases): Use first author plus 'et al.'

\Copyright{L\'aszl\'o Kozma}%TODO mandatory, please use full first names. LIPIcs license is "CC-BY";  http://creativecommons.org/licenses/by/3.0/


%TODO mandatory; please add comma-separated list of keywords

\category{}%optional, e.g. invited paper

\relatedversion{}%optional, e.g. full version hosted on arXiv, HAL, or other respository/website
\supplement{}%optional, e.g. related research data, source code, ... hosted on a repository like zenodo, figshare, GitHub, ...

\acknowledgements{}%optional

\nolinenumbers %uncomment to disable line numbering

\hideLIPIcs  %uncomment to remove references to LIPIcs series (logo, DOI, ...), e.g. when preparing a pre-final version to be uploaded to arXiv or another public repository

\EventEditors{John Q. Open and Joan R. Access}
\EventNoEds{2}
\EventLongTitle{42nd Conference on Very Important Topics (CVIT 2016)}
\EventShortTitle{CVIT 2016}
\EventAcronym{CVIT}
\EventYear{2016}
\EventDate{December 24--27, 2016}
\EventLocation{Little Whinging, United Kingdom}
\EventLogo{}
\SeriesVolume{42}
\ArticleNo{23}
\begin{document}

\maketitle

%TODO mandatory: add short abstract of the document
\begin{abstract}
Permutation patterns and pattern avoidance have been intensively studied in combinatorics and computer science, going back at least to the seminal work of Knuth on stack-sorting (1968). Perhaps the most natural algorithmic question in this area is deciding whether a given permutation of length $n$ contains a given pattern of length $k$. 

In this work we give two new algorithms for this well-studied problem, one whose running time is $n^{0.44k + o(k)}$, and one whose running time is the better of $O(1.6181^n)$ and $n^{k/2 + o(k)}$. These results improve the earlier best bounds of Ahal and Rabinovich (2000), and Bruner and Lackner (2012), and are the fastest algorithms for the problem when $k = \Omega(\log{n})$. When $k=o(\log{n})$, the parameterized algorithm of Guillemot and Marx (2013) dominates.

Our second algorithm uses polynomial space and is significantly simpler than all previous approaches with comparable running times, including an $n^{k/2 + o(k)}$ algorithm proposed by Guillemot and Marx. Our approach can be summarized as follows: ``for every matching of the even-valued entries of the pattern, try to match all odd-valued entries left-to-right''. For the special case of patterns that are \emph{Jordan-permutations}, we show an improved, subexponential running time. 

\end{abstract}

\section{Introduction}
\label{sec:intro}

Let $[n] = \{1,\dots,n\}$. Given two permutations $t:[n] \rightarrow [n]$, % = (t_1, \dots, t_n)$ 
and $\pi:[k] \rightarrow [k]$,
%$\pi = (\pi_1, \dots, \pi_k)$, 
we say that $t$ \emph{contains} $\pi$, if there are indices $1 \leq i_1 < \cdots < i_k \leq n$ such that $t({i_j}) < t({i_\ell})$ if and only if $\pi({j}) < \pi(\ell)$, for all $1 \leq j,\ell \leq k$. In other words, $t$ contains $\pi$, if the sequence $(t(1), \dots, t(n))$ has a (possibly non-contiguous) subsequence with the same ordering as $(\pi(1),\dots,\pi(k))$, otherwise $t$ \emph{avoids} $\pi$. For example, $t = (1,5,4,6,3,7,8,2)$ contains $(2,3,1)$, because its subsequence $(5,6,3)$ has the same ordering as $(2,3,1)$; on the other hand, $t$ avoids $(3,1,2)$. 

Knuth showed in 1968~\cite[§\,2.2.1]{knuth68}, that permutations sortable by a single stack are exactly those that avoid $(2,3,1)$. Sorting by restricted devices has remained an active research topic~\cite{TarjanSorting, Pratt, Rosenstiehl, bona2003survey,AlbertB15, Pantone}, but permutation pattern avoidance has taken on a life of its own (especially after the influential work of Simion and Schmidt~\cite{Simion}), becoming an important subfield of combinatorics. For more background on permutation patterns and pattern avoidance we refer to the extensive survey~\cite{vatter2014} and relevant textbooks~\cite{bona1, bona2, kitaev}. 

Perhaps the most important enumerative result related to permutation patterns is the Stanley-Wilf conjecture, raised in the late 1980s and proved in 2004 by Marcus and Tardos~\cite{MT}.  It states that the number of length-$n$ permutations which avoid a fixed pattern $\pi$ is bounded by ${c(\pi)}^n$, where $c(\pi)$ is a quantity independent of $n$. (Marcus and Tardos proved the result in the context of $0/1$ matrices, answering a question of F\"uredi and Hajnal~\cite{FurediH92}, which was shown by Klazar~\cite{klazar2000furedi} to imply the Stanley-Wilf conjecture.) 

A fundamental algorithmic problem in this context is \emph{Permutation Pattern Matching} (PPM): 
Given a length-$n$ permutation $t$ (``text'') and a length-$k$ permutation $\pi$ (``pattern''), decide whether $t$ contains $\pi$. 

Solving PPM is a bottleneck in experimental work on permutation patterns~\cite{Albert_algo}. The problem and its variants also arise in practical applications, e.g.\ in computational biology~\cite[§\,2.4]{kitaev} and time-series analysis~\cite{timeseries1, timeseries2, timeseries3}. Unfortunately, in general, PPM is $\NP$-complete, as shown by Bose, Buss, and Lubiw~\cite{BBL} in 1998. (This is in contrast to e.g.\ string matching problems that are solvable in polynomial time.) 

An obvious algorithm for PPM is to enumerate all ${n \choose k}$ length-$k$ subsequences of $t$, and check whether any of them has the same ordering as $\pi$. 
% ${n^k}/{k!}$ or $2^n/\sqrt{n}$. Checking whether a given subsequence matches the pattern can be done in linear time.  
The first result to break this ``triviality barrier'' was the $n^{2k/3+o(k)}$-time algorithm of Albert, Aldred, Atkinson, and Holton~\cite{Albert_algo}. Around the same time, Ahal and Rabinovich~\cite{Ahal} obtained the running time $n^{0.47k+o(k)}$. The two algorithms are based on a similar dynamic programming approach, but they differ in essential details.

Informally, in both algorithms, the entries of the pattern $\pi$ are matched one-by-one to entries of the text $t$, observing the order-restrictions imposed by the current partial matching. The key observation is that only a subset of the matched entries need to be remembered, namely those that form a certain ``boundary'' of the partial matching. The maximum size of this boundary depends on the matching strategy used, and the best attainable value is a graph-theoretic parameter of a certain graph constructed from the pattern (the parameter is essentially the \emph{pathwidth} of the \emph{incidence graph} of $\pi$). We review this framework in §\,\ref{sec:dynalgo}.

In the algorithm of Albert et al.\ the pattern-entries are matched in the simplest, left-to-right order. In the algorithm of Ahal and Rabinovich, the pattern-entries are matched in a uniform random order, interspersed with greedy steps that reduce the boundary. (For the purely random strategy, they show a weaker $n^{0.54k + o(k)}$ bound, and they describe further heuristics, without analysis.) 

Our first result is a new algorithm for PPM (Algorithm~M), improving the bounds of~\cite{Albert_algo, Ahal}. 

\begin{theorem}\label{thm1}
Algorithm~M solves Permutation Pattern Matching in time $n^{0.44k + o(k)}$.
\end{theorem}

The algorithm uses dynamic programming, but selects the pattern-entries to be matched using an optimized, \emph{global} strategy, differing significantly from the previous approaches. In addition to being the first (admittedly small) improvement in a long time on the complexity of PPM for large patterns, our approach is deterministic, and has the advantage of a more transparent analysis. (The analysis of the random walk in~\cite{Ahal} is based on advanced probabilistic arguments, and appears in the journal paper only as a proof sketch.)

\medskip

In 2013, Guillemot and Marx~\cite{GM} obtained the breakthrough result of a PPM algorithm with running time $2^{O(k^2 \log{k})} \cdot n$. This result established the \emph{fixed-parameter tractability} of the problem in terms of the pattern length. Their algorithm builds upon the Marcus-Tardos proof of the Stanley-Wilf conjecture and introduces a novel decomposition of permutations. For the (arguably most natural) case of constant-size patterns, the Guillemot-Marx algorithm has linear running time. (Due to the large constants involved, it is however, not clear how efficient it is in practice.) Subsequently, Fox~\cite{JFox} refined the Marcus-Tardos result, thereby improving the Guillemot-Marx bound, removing the $\log{k}$ factor from the exponent. Whether the dependence on $k$ can be further improved remains an intriguing open question.

In light of the Guillemot-Marx result, for constant-length patterns, the PPM problem is well-understood. However, for patterns of length e.g.\ $k \approx \log_2{n}$ or larger, the complexity of the problem is open, and in this regime, Theorem~\ref{thm1} is an improvement over previous results. 

Guillemot and Marx also describe an alternative, \emph{polynomial-space} algorithm, with running time $n^{k/2+o(k)}$, i.e.\ slightly above the Ahal-Rabinovich bound~\cite[§\,7]{GM}. Although simpler than their main result, this method is still rather complex---it works by decomposing the text into $2\lceil \sqrt{n} \rceil$ monotone subsequences (such a decomposition exists by the Erd\H{o}s-Szekeres theorem), and solving as a subroutine, a certain constraint-satisfaction problem whose tractability is implied by a nontrivial structural property. Our second result (Algorithm~S) matches these time and space bounds by an exceedingly simple approach.

Expressed in terms of $n$ only, none of the mentioned running times improve, in the worst case, upon the trivial $2^n$. (Consider the case of a pattern of length $k = \Omega( n/\log{n})$.) The first non-trivial bound in this parameter range was obtained by Bruner and Lackner~\cite{BL}; their algorithm runs in time $O(1.79^n)$.   

The algorithm of Bruner and Lackner works by decomposing both the text and the pattern into \emph{alternating runs} (consecutive sequences of increasing or decreasing elements). They then use this decomposition to restrict the space of admissible matchings. The exponent in the running time is, in fact, the \emph{number of runs} of $T$, which can be as large as $n$. The approach is compelling and intuitive, the details, however, are intricate (the description of the algorithm and its analysis in~\cite{BL} take over 24 pages).

Our second result also improves this running time, with an algorithm that can be described and analysed in a few paragraphs. 

\begin{theorem}\label{thm2}
Algorithm~S solves Permutation Pattern Matching using polynomial space, in time $n^{k/2 + o(k)}$ or $O(1.6181^n)$. 
\end{theorem}

At the heart of Algorithm~S is the following observation: if all even-valued entries of the pattern $\pi$ are matched to entries of the text $t$, then verifying whether the remaining (odd) entries of $\pi$ can be correctly matched takes only a linear-time, left-to-right sweep through both $\pi$ and $t$. 

\medskip

Beyond the general case, the PPM problem has been extensively studied when either $t$ or $\pi$ come from some restricted family of permutations. Examples include \emph{separable} permutations~\cite{BBL, Ibarra, Saxena, Albert_algo}, i.e.\ avoiding $(2,4,1,3)$ and $(3,1,4,2)$, $k$-increasing or $k$-decreasing patterns~\cite{Chang}, $k$-monotone text~\cite{GM, vialette}, patterns of length $3$ or $4$~\cite{Albert_algo}, patterns avoiding $(3,2,1)$~\cite{321}, text or pattern avoiding $(2,1,3)$ and $(2,3,1)$~\cite{231}, \emph{linear} permutations~\cite{Ahal}, permutations with few runs~\cite{BL}. 

In a similar vein, we consider the case of \emph{Jordan-permutations}, a natural family of geometrically-defined permutations with applications in intersection problems of computational geometry~\cite{rosenstiehl:hal-00259765}. Jordan-permutations were studied by Hoffmann, Mehlhorn, Rosenstiehl, and Tarjan~\cite{hoffmann1986sorting}, who showed that they can be sorted with a linear number of comparisons, using level-linked trees. A Jordan permutation is generated by the intersection-pattern of two simple curves in the plane. Label the intersection points between the curves in increasing order along the first curve, and read out the labels along the second curve; the obtained sequence is a Jordan-permutation (Figure~\ref{fig1}). We show that if the pattern is a Jordan-permutation, then PPM can be solved in sub-exponential time. 

\begin{theorem}\label{thm4}
If $\pi$ is a Jordan permutation, then PPM can be solved in time $n^{O(\sqrt{k})}$.
\end{theorem}

The improvement comes from the observation that the incidence graph of Jordan-permutations is (by construction) planar, allowing the use of planar separators in generating a good matching order in the dynamic programming framework of Algorithm~M.

This observation suggests a more general question. Which permutations have incidence graphs with sublinear-size separators?  We observe that many of the special families of patterns considered in the literature can be understood in terms of avoiding certain \emph{fixed} patterns (that is, the pattern $\pi$ avoids some smaller pattern $\sigma$). The following conjecture thus appears natural, as it would unify and generalize a number of results in the literature.

\begin{conjecture}\label{conj}
If the pattern $\pi$ avoids an arbitrary pattern $\sigma$ of length $O(1)$, then PPM (with text $t$ and pattern $\pi$) can be solved in time $2^{o(n)}$.
\end{conjecture}

Conjecture~\ref{conj} appears plausible, especially in light of the extremal results of Fox~\cite{JFox} related to the Stanley-Wilf conjecture. Characterizing the base $c(\pi)$ of the Stanley-Wilf bound remains a deep open question. Fox has recently shown that, contrary to prior conjectures, $c(\pi)$ is exponential in $|\pi|$ for most patterns $\pi$. In some special cases it is known that $c(\pi)$ is subexponential, and Fox conjectures~\cite[Conj.~1]{JFox} that this is the case whenever the pattern $\pi$ is itself pattern-avoiding; Conjecture~\ref{conj} can be seen as an algorithmic counterpart of this question. 

A possible route towards proving Conjecture~\ref{conj} is to show that avoidance of a length-$d$ pattern in $\pi$ (and the resulting $f(d)$-wide decomposition given by Guillemot and Marx) implies the avoidance of some $g(d)$-size minor in the incidence graph $G_\pi$. We only remark here, that one cannot hope for a forbidden \emph{grid minor}, as there are $O(n)$-permutations that avoid an $O(1)$-length pattern, yet contain a $\sqrt{n} \times \sqrt{n}$ grid in their incidence graph (§\,\ref{sec:spec}). 

\subparagraph*{Further related work.} 

Only \emph{classical patterns} are considered in this paper; variants in the literature include \emph{vincular}, \emph{bivincular}, \emph{consecutive}, and \emph{mesh} patterns; we refer to~\cite{BLcomp} for a survey of related computational questions.

Newman et al.\ \cite{Newman} study pattern matching in a \emph{property-testing} framework (aiming to distinguish pattern-avoiding sequences from those that contain \emph{many copies} of the pattern). In this setting, the focus is on the \emph{query complexity} of different approaches, and sampling techniques are often used; see also~\cite{BenEliezerC, foxPT}.

A different line of work investigates whether standard algorithmic problems on permutations (e.g.\ sorting, selection) become easier if the input can be assumed to be pattern-avoiding~\cite{Arthur, FOCS15}.

\subparagraph*{Structure of the paper.}

In §\,\ref{sec:prel} we introduce the concepts necessary to state and prove our results. In §\,\ref{sec:algo} we describe and analyse our two algorithms; in §\,\ref{sec:algo1} the simpler Algorithm~S, and in §\,\ref{sec:algo2} the improved Algorithm~M, thereby proving Theorems~\ref{thm1} and \ref{thm2}. We describe the dynamic programming framework used by Algorithm~M in §\,\ref{sec:dynalgo}. In §\,\ref{sec:spec} we discuss the special cases (Theorem~\ref{thm4} and Conjecture~\ref{conj}), and in §\,\ref{sec:open} we conclude with further questions.

\section{Preliminaries}\label{sec:prel}

A length-$n$ permutation $\sigma$ is a bijective function $\sigma: [n] \rightarrow [n]$, alternatively viewed as the sequence $(\sigma(1), \dots, \sigma(n))$. Given a length-$n$ permutation $\sigma$, we denote as $S_\sigma = \{(i,\sigma(i)) \mid 1 \leq i \leq n\}$ the \emph{set of points} corresponding to permutation $\sigma$. 

For a point $p \in S_\sigma$ we denote its first entry as $p.x$, and its second entry as $p.y$, referring to these values as the \emph{index}, respectively, the \emph{value} of $p$. Observe that for every $i \in [n]$, we have $|\{p\in S_\sigma \mid p.x = i\}| = |\{p\in S_\sigma \mid p.y = i\}| = 1$.

We define four neighbors of a point $(x,y) \in S_\sigma$ as follows.
\begin{eqnarray*}
N^R((x,y)) & = & (x+1,~ \sigma(x+1)), \\
N^L((x,y)) & = & (x-1,~ \sigma(x-1)), \\
N^U((x,y)) & = & (\sigma^{-1}(y+1),~ y+1), \\
N^D((x,y)) & = & (\sigma^{-1}(y-1),~ y-1). 
\end{eqnarray*}

The superscripts $R$, $L$, $U$, $D$ are meant to evoke the directions \emph{right}, \emph{left}, \emph{up}, \emph{down}, when plotting $S_\sigma$ in the plane. Some neighbors of a point may coincide. When some index is out of bounds, we let the offending neighbor be a ``virtual point'' as follows: $N^R(n,i) = N^U(i,n) = (\infty,\infty)$, and $N^L(1,i) = N^D(i,1) = (0,0)$, for all $i \in [n]$. The virtual points are not contained in $S_\sigma$, we only define them to simplify some of the statements.

\begin{figure*}[h]
  \centering
  \includegraphics[width=12cm]{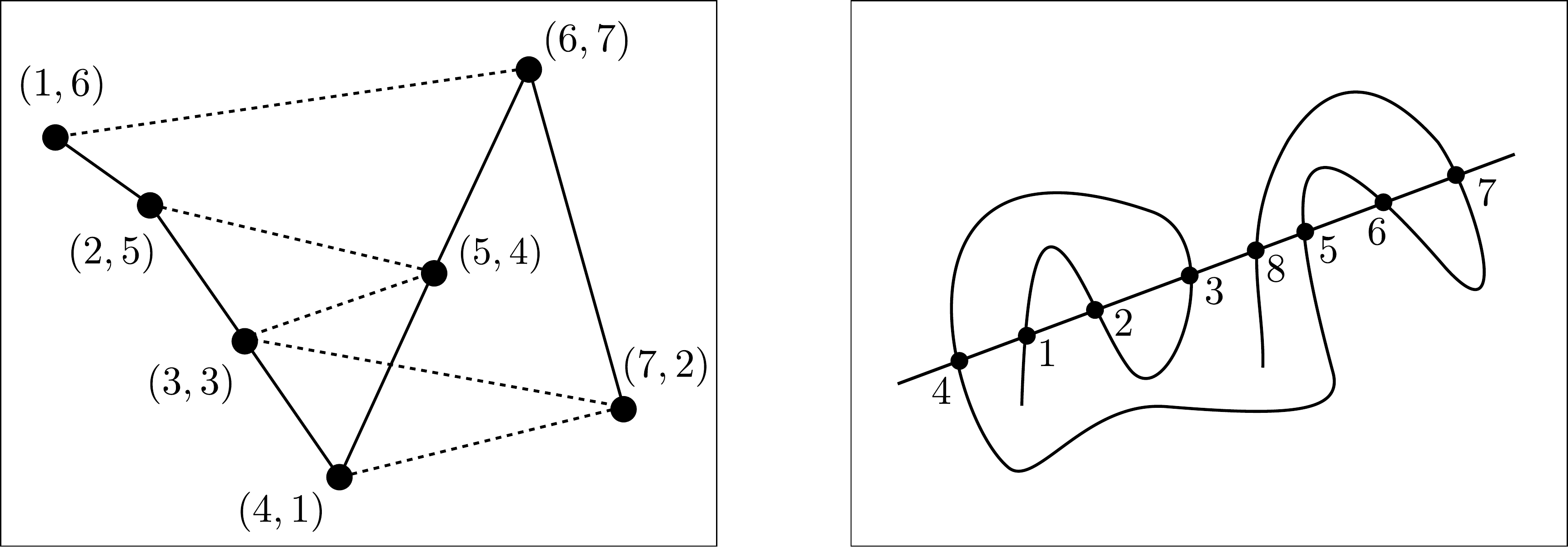}
  \caption{(left) Permutation $\pi = (6,5,3,1,4,7,2)$ and its incidence graph $G_\pi$. Solid lines indicate neighbors by index, dashed lines indicate neighbors by value (lines may overlap). Indices plotted on $x$-coordinate, values plotted on $y$-coordinate. (right) Jordan-permutation $(4,1,2,3,8,5,6,7)$\label{fig1}}
\end{figure*}

The \emph{incidence graph} of a permutation $\sigma$ is $G_\sigma = (S_\sigma,E_\sigma)$, where $$E_\sigma = \left\{\left(p,N^{\upalpha}(p)\right) \mid \upalpha \in \{R,L,U,D\}, \mbox{~and~~} p, N^{\upalpha}(p) \in S_\sigma\right\}.$$ In words, each point is connected to its (at most) four neighbors: its successor and predecessor by index, and its successor and predecessor by value. It is easy to see that $G_\sigma$ is a union of two Hamiltonian paths on the same set of vertices, and that this is, in fact, an exact characterization of permutation incidence-graphs. (See Figure~\ref{fig1} for an illustration.)

Throughout the paper we consider a text permutation $t: [n] \rightarrow [n]$, and a pattern permutation $\pi : [k] \rightarrow [k]$, where $n \geq k$. We give an alternative definition of the Permutation Pattern Matching (PPM) problem in terms of embedding $S_\pi$ into $S_t$. 

Consider a function $f:S_\pi \rightarrow S_t$. We say that $f$ is a \emph{valid embedding} of $S_\pi$ into $S_t$ if for all $p \in S_\pi$ the following hold:
\begin{eqnarray}
f(N^L(p)).x & < ~~f(p).x  & < ~~f(N^R(p)).x, \mbox{~~and}\\
f(N^D(p)).y & < ~~f(p).y & < ~~f(N^U(p)).y, 
\end{eqnarray}
whenever the corresponding neighbor $N^{\upalpha}(p)$ is also in $S_\pi$, i.e.\ not a virtual point. In words, valid embeddings preserve the relative positions of neighbors in the incidence graph.

\begin{lemma}\label{eqlem}
Permutation $t$ contains permutation $\pi$ if and only if there exists a valid embedding $f:S_\pi \rightarrow S_t$.
\end{lemma}

For sets $A \subseteq B \subseteq S_\pi$ and functions $g : A \rightarrow S_t$ and $f: B \rightarrow S_t$ we say that $g$ is the \emph{restriction} of $f$ to $A$, denoted $g = f |_{A}$, if  $g(i) = f(i)$ for all $i \in A$. In this case, we also say that $f$ is the \emph{extension} of $g$ to $B$. 
Restrictions of valid embeddings will be called \emph{partial embeddings}. We observe that if $f : B \rightarrow S_t$ is a partial embedding, then it satisfies conditions (1) and (2) with respect to all edges in the induced graph $G_\pi[B]$, i.e.\ the corresponding inequality holds whenever $p, N^\upalpha(p) \in B$.

\section{Algorithms for pattern matching} \label{sec:algo}

We start in §\,\ref{sec:algo1} with the simpler Algorithm~S, proving Theorem~\ref{thm2}. In §\,\ref{sec:dynalgo} we describe the dynamic programming framework used in Algorithm~M (Algorithm~S does not require this). In §\,\ref{sec:algo2} we describe and analyse Algorithm~M, proving Theorem~\ref{thm1}.
 
\subsection{The even-odd method} \label{sec:algo1}

Let $(Q^E, Q^O)$ be the partition of $S_\pi$ into points with even and odd indices. Formally, $Q^{E} = \{(2k, \pi(2k)) \mid 1 \leq k \leq \lfloor k/2 \rfloor \}$, and $Q^{O} = \{(2k-1,\pi(2k-1)) \mid 1 \leq k \leq \lceil k/2 \rceil \}$. 

Suppose $t$ contains $\pi$. Then, by Lemma~\ref{eqlem}, there exists a valid embedding $f:S_\pi \rightarrow S_t$. 
We start by \emph{guessing} a partial embedding $g_0: Q^{E} \rightarrow S_\pi$. (For example, $g_0 = f|_{Q^E}$ is such a partial embedding.)
We then extend $g_{0}$ step-by-step, adding points to its domain, until it becomes a valid embedding $S_\pi \rightarrow S_t$. 

Let $p_1, \dots, p_{\lceil k/2 \rceil}$ be the elements of $Q^{O}$ in increasing order of \emph{value}, i.e.\ $1 \leq p_1.y < \cdots < p_{\lceil k/2 \rceil}.y \leq n$, and let $P_0 = \emptyset$ and $P_i = P_{i-1} \cup \{p_i\}$, for $1 \leq i \leq \lceil k/2 \rceil$. For all $i$, we maintain the invariant that $g_i$ is a restriction of some valid embedding to $Q^{E} \cup P_i$. By our choice of $g_0$, this is true initially for $i=0$.

In the $i$-th step (for $i=0,\dots,\lceil k/2 \rceil-1$), we extend $g_i$ to $g_{i+1}$ by mapping the next point $p_{i+1}$ onto a suitable point in $S_t$. For $g_{i+1}$ to be a restriction of a valid embedding, it must satisfy conditions (1) and (2) on the relative position of neighbors. Observe that all, except possibly one, of the neighbors of $p_{i+1}$ are already embedded by $g_i$. This is because $N^L(p_{i+1})$ and $N^R(p_{i+1})$ have even index, are thus in $Q^E$, unless they are virtual points and thus implicitly embedded. The point $N^D(p_{i+1})$ is either an even-index point, and thus in $Q^E$, or the virtual point $(0,0)$ and thus implicitly embedded, or an odd-index point, in which case, by our ordering, it must be $p_i$, and thus, contained in $P_i$. The only neighbor of $p_{i+1}$ possibly not embedded is $N^U(p_{i+1})$. 

If we map $p_{i+1}$ to a point $q \in S_t$, we have to observe the constraints $g_{i}(N^{L}(p_{i+1})).x < q.x < g_{i}(N^{R}(p_{i+1})).x$, and $g_{i}(N^{D}(p_{i+1})).y < q.y$. If $N^U(p_{i+1})$ is also in the domain of $g_i$, then we have the additional constraint $q.y < g_{i}(N^{U}(p_{i+1})).y$.

These constraints determine an (open) axis-parallel box, possibly extending upwards infinitely (in case only three of the four neighbors of $p_{i+1}$ are embedded so far). Assuming $g_i$ is a restriction of a valid embedding $f$, the point $f(p_{i+1})$ must satisfy all constraints, it is thus contained in this box. We extend $g_i$ to obtain $g_{i+1}$ by mapping $p_{i+1}$ to a point $q \in S_t$ in the constraint-box, and if there are multiple such points, we pick the one that is \emph{lowest}, i.e.\ the one with smallest value $q.y$. 

The crucial observation is that if $g_i$ is a partial embedding, then $g_{i+1}$ is also a partial embedding, and the correctness of the procedure follows by induction.

Indeed, some valid embedding $f' : S_\pi \rightarrow S_t$ must be the extension of $g_{i+1}$. If $q = f(p_{i+1})$, then $f'$ is $f$ itself. Otherwise, let $f'$ be identical with $f$, except for mapping $p_{i+1} \rightarrow q$ (instead of mapping $p_{i+1} \rightarrow f(p_{i+1})$). The only conditions of a valid embedding that may become violated are those involving $p_{i+1}$. The conditions $f'(N^L(p_{i+1})).x < f'(p_{i+1}).x < f'(N^R(p_{i+1})).x$ and $f'(N^D(p_{i+1})).y < f'(p_{i+1}).y$ hold by our choice of $q$.

\begin{figure*}[h]
  \centering
  \includegraphics[width=12cm]{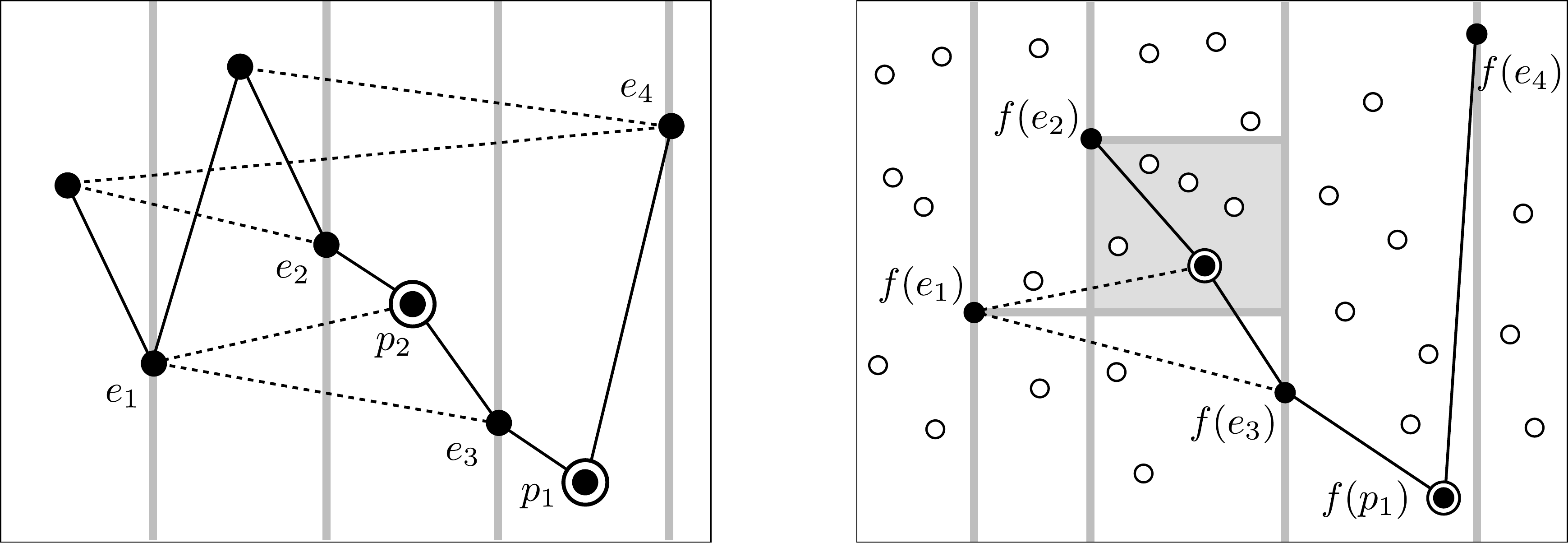}
  \caption{(left) Pattern $\pi = (6,3,8,5,4,2,1,7)$ and its incidence graph $G_\pi$. Solid lines indicate neighbors by index, dashed lines indicate neighbors by value (lines may overlap). (right) Text permutation $t$, points shown as circles. Partial embedding of $\pi$ shown with filled circles. Vertical bars mark even-index points $e_1$, $e_2$, $e_3$, $e_4$. Double circles mark the first two odd-index points $p_1$, $p_2$. Shaded box indicates constraints for embedding $p_{2}$, determined by $N^U(p_{2}) = N^L(p_2) = e_2$, $N^D(p_{2}) = e_1$, and $N^R(p_{2}) = e_3$. Observe that $p_2$ is mapped to lowest point (by value) that satisfies constraints. Revealed edges of $G_\pi$ are shown.  \label{algo}}
\end{figure*}

The condition $f'(p_{i+1}).y < f'(N^U(p_{i+1})).y$ holds a fortiori since we picked the \emph{lowest point} in a box that also contained $f(p_{i+1})$, in other words, $f'(p_{i+1}).y = q.y \leq f(p_{i+1}).y < f(N^U(p_{i+1})).y = f'(N^U(p_{i+1})).y$. Thus, $g_{i+1}$ is a partial embedding, which concludes the argument. See Figure~\ref{algo} for illustration.

Assuming that our initial guess $g_{0}$ was correct, we succeed in constructing a valid embedding that certifies the fact that $t$ contains $\pi$. We remark that guessing $g_0$ should be understood as trying all possible embeddings of $Q^E$. If our choice of $g_0$ is incorrect, i.e.\ not a partial embedding, then we reach a situation where we cannot extend $g_i$, and we abandon the choice of $g_0$. If extending $g_0$ to a valid embedding fails for all initial choices, we conclude that $t$ does not contain $\pi$. The resulting Algorithm~S is described in Figure~\ref{algoS}. 

\begin{figure}
\begin{center}
\begin{algorithm}[H]
\DontPrintSemicolon
\NoCaptionOfAlgo
\caption{Algorithm~S:}

\For{\textbf{\emph{all~}} $g_0: Q^E \rightarrow S_t$}{
\textbf{if} $g_0$ not valid, \textbf{next} $g_0$ \;
\For{$i\gets 0$ \KwTo $ \lceil k/2 \rceil - 1$}{
    \textbf{let} $q \in S_t$ with minimum $q.y$ such that: \;
    ~~~~$g_i(N^L(p_{i+1})).x < q.x < g_i(N^R(p_{i+1})).x$ \; 
    ~~~~$g_i(N^D(p_{i+1})).y < q.y$ \; 
    ~~~~$g_i(N^U(p_{i+1})).y> q.y$ ~~~(in case $N^U(p_{i+1}) \in Q^E$) \;
\textbf{if} no such $q$, \textbf{next} $g_0$ \;
\textbf{extend} $g_i$ to $g_{i+1}$ by mapping $p_{i+1} \rightarrow q$ \;
}
\textbf{return} $g_{\lceil k/2 \rceil}$
}
\textbf{return} ``$t$ avoids $\pi$''
\end{algorithm}
\caption{\label{algoS}Finding a valid embedding of $S_\pi$ into $S_t$, or reporting that $t$ avoids $\pi$, with precomputed $Q^E$ (even-index points of $S_\pi$) and $(p_1, \dots, p_{\lceil k/2 \rceil})$ (odd-index points of $S_\pi$ sorted by value). \;}
\end{center}
\end{figure}

The space requirement is linear in the input size; apart from minor bookkeeping, only a single embedding must be stored at all times.

To analyse the running time, observe first, that $g_0$ must map points in $Q^E$ to points in $S_t$, preserving their left-to-right order (by index), and their bottom-to-top order (by value). The first condition can be enforced directly, by considering only \emph{subsequences} of $t$. This leads to ${n \choose \lfloor k/2 \rfloor}$ choices for $g_0$ in the outer loop. The second condition can be verified in a linear time traversal of $G_\pi$ (this is the second line of Algorithm~S). 

All remaining steps can be  performed using straightforward data structuring: we need to traverse to neighbors in the incidence graph, to go from $x$ to $g_i(x)$ and back, and to answer rectangle-minimum queries; all can be achieved in constant time, with a polynomial time preprocessing. We can in fact do away with rectangle queries, since candidate points of $t$ are considered in increasing order of value---the inner loop thus consists of a single sweep through $\pi$ and $t$, which can be implemented in $O(n)$ time. By a standard bound on the binomial coefficient, the claimed running time of $n^{k/2 + o(k)}$ follows.

We refine the analysis, observing that in the outer loop only those embeddings $g_0$ (i.e.\ subsequences of $t$) need to be considered, that leave a gap of at least one point between each successive entry (to allow for embedding the odd-index points). The number of subsequences with this property is ${{n-k/2} \choose {k/2}}$; first embed all even-index entries with a minimum required gap of one between them, then distribute the remaining total gap of $n-k$ among the $ k/2 $ slots. 
To bound this quantity, denote $\upalpha = \frac{k}{2n}$ and $m = n - \upalpha{n}$, 
to obtain ${m \choose {m \upbeta}}$, where $\upbeta = \frac{\upalpha}{1 - \upalpha}$.
A standard upper bound for this quantity (see e.g.~\cite[§\,11]{Cover_Thomas}) is $2^{m \cdot H(\upbeta)}$, where $H$ is the binary entropy function $H(x) = -\log_2{(x^x \cdot (1-x)^{1-x})}$. 

Our upper bound is thus $2^{n \cdot {(1-\upalpha)} \cdot H(\upalpha/(1-\upalpha))}$. After simplification, we obtain the expression $[B(\upalpha)]^n$, where $$B (\upalpha) = \frac{(1-\upalpha)^{1-\upalpha}}{\upalpha^\upalpha \cdot (1-2\upalpha)^{(1-2\upalpha)}}.$$

In the range of interest $0 < \upalpha < 0.5$, we find $B(\upalpha)$ to be maximized for $\upalpha = \frac{1}{2} - \frac{1}{2\sqrt{5}}$, attaining a value smaller than $1.6181$. We obtain thus the upper bound $O(1.6181^n)$ for the running time.

The efficient enumeration of initial embeddings with the required property can be done with standard techniques, see e.g.~\cite{comb_book}. Note that the algorithm can equivalently be implemented in the variant where both $t$ and $\pi$ are transposed, i.e.\ by embedding even \emph{values} first, followed by odd values sorted by \emph{index}, as described in~§\,\ref{sec:intro}. 

Finally, we remark that instead of trying all embeddings $g_0$, it may be more practical to build such an embedding incrementally, using backtracking. This allows the process to ``fail early'' if a certain embedding can not be extended to any partial embedding of $Q^E$.  
The order in which points of $Q^E$ are considered in the backtracking process can affect the performance significantly, see~\cite{knuth1975estim,knuth2000dancing} for consideration of similar issues. Alternatively, a modification of the dynamic programming approach of §\,\ref{sec:dynalgo} and §\,\ref{sec:algo2} may also be used to enumerate all valid initial $g_0$.

\subsection{Dynamic programming approach}\label{sec:dynalgo}
We review the dynamic programming framework that Algorithm~M (§\,\ref{sec:algo2}) shares with the previous algorithms of Albert et al.\ \cite{Albert_algo} and Ahal and Rabinovich~\cite{Ahal}. We refer to these works for a more detailed exposition.

The idea is to fix an \emph{embedding order} $\tau$ in which the elements of $S_\pi$ are processed. Let $\tau : [k] \rightarrow [k]$ be a permutation, and let $\emptyset = P_0 \subset P_1 \subset \cdots \subset P_k = S_\pi$, where $P_i =  P_{i-1} \cup \{p_i\}$, and $p_i = (\tau(i),\pi(\tau(i)))$.

For $i = 1, \dots, k$, we find embeddings $g_i : P_i \rightarrow S_t$ that extend the previously found embeddings $g_{i-1}$, by mapping $p_i$ to a suitable target $q \in S_t$. The difference from Algorithm~S is that we consider \emph{all possible} targets (that satisfy the neighbor-constraints with respect to already mapped neighbors), and we store all (so far) valid embeddings $g_i$ in a table.

More precisely, we store all embeddings of $P_i$ that do not violate any neighborhood-constraint in $G_\pi[P_i]$. In the $i$-th step, for all stored $g_i$ embeddings, we find all possible extensions, mapping $p_i$ to $q$ such that $g_i(N^L(p_i)).x < q.x <  g_i(N^R(p_i)).x$, and $g_i(N^D(p_i)).y < q.y <  g_i(N^U(p_i)).y$, whenever the respective neighbor of $q$ is in the domain of $g_i$, i.e.\ already embedded by $g_i$. 

The key to improving this basic approach is the observation that for each embedding $g_i$ it is sufficient to store those points that have neighbors in $G_\pi$ that are \emph{not yet embedded}. (Points whose neighbors are all embedded cannot influence future choices.) We thus define, for a set $P \subseteq S_\pi$ the \emph{boundary} set $$\bd(P) = \{q \mid q \in P, ~~ N^{\upalpha}(q) \in S_\pi \setminus P, \mbox{~~for some~} \upalpha \in \{R,L,U,D\} \}.$$

Instead of storing the embeddings $g_i : P_i \rightarrow S_t$, we only store their restrictions $g_i|_{\bd{(P_i)}}$. (As different embeddings may have the same restriction, careful data structuring is required to prune out duplicates; see~\cite{Ahal}.)

The total space- and time-requirement of the resulting algorithm is dominated by the number of \emph{essentially different} embeddings $g_i$ (i.e.\ those with different restrictions to the boundary). At a given step $i$, the number of possible boundary-embeddings is at most ${n \choose |\bd(P_i)|}$. Observe that this quantity depends on the sets $P_i$, which are in turn determined by the embedding order $\tau$. Let us therefore define $\bd_\tau(\pi) = \max_i {|\bd(P_i)|}$. 

The quantity $\min_\tau \bd_\tau(\pi)$ is known in the literature as the \emph{vertex-separation number}, computed here for the graph $G_\pi$. For an arbitrary graph, this quantity equals the \emph{pathwidth} of the graph~\cite{NK,DowneyFellows}. An upper bound $\upalpha \cdot k$ on the pathwidth of $G_\pi$ thus implies an algorithm with running time at most $n^{\upalpha \cdot k + o(k)}$ (by a standard upper bound on the binomial coefficient), and the problem reduces to finding an embedding order $\tau$ with small $\bd_\tau(\pi)$.

Albert et al.\ show that if $\tau$ is the identity permutation, then $\bd_\tau(\pi) \leq 2k/3+1$. Ahal and Rabinovich show that the $2k/3+1$ bound cannot be improved for any fixed $\tau$ (i.e.\ $\tau$ independent of $\pi$), and describe a randomized construction of $\tau$ with $\bd_\tau(\pi) \leq 0.47 k + o(k)$. They further show that there are permutations $\pi$ for which $\bd_\tau(\pi) > 0.036 k$, with arbitrary $\tau$. (This follows from a result of Bollob\'as on random $4$-regular graphs~\cite{bollobas}.) 

We remark that Algorithm~S (§\,\ref{sec:algo1}) can also be seen as an embedding order $\tau$, with $\bd_\tau(\pi) \leq k/2 + o(k)$, it can thus be easily adapted to the dynamic programming framework with similar running time, although at the cost of exponential space. (The key to the efficiency of Algorithm~S is that we need not store more than one embedding.)

\subsection{Improved dynamic programming} \label{sec:algo2}

As in §\,\ref{sec:algo1}, let $Q^E$ and $Q^O$ denote the even-index, resp.\ odd-index points of $S_\pi$.
Let $s = \lfloor \log_2{k} \rfloor$, and let $I_j = [(j-1) \cdot k/(2s) + 1, ~~j \cdot k/(2s)]$, for $j=1, \dots, 2s$.  In words, partition $[k]$ into equal-length contiguous intervals $I_1, \dots, I_{2s}$. Assume for simplicity that $2s$ divides $k$; we can ensure this by appropriate padding of $\pi$ (after fixing $s$). Let $P_j = \{p \in S_\pi \mid p.y \in I_j \}$, i.e.\ the points of $S_\pi$ whose value falls in the $j$-th interval. 

We describe the embedding order $\tau$ in three stages, and show that $\bd_\tau(\pi) 
 \leq 0.4375 k + o(k)$, which, by the discussion in §\,\ref{sec:dynalgo} yields the running time claimed in Theorem~\ref{thm1}. First we summarize the process. 

\begin{figure*}[h]
  \centering
  \includegraphics[width=9cm]{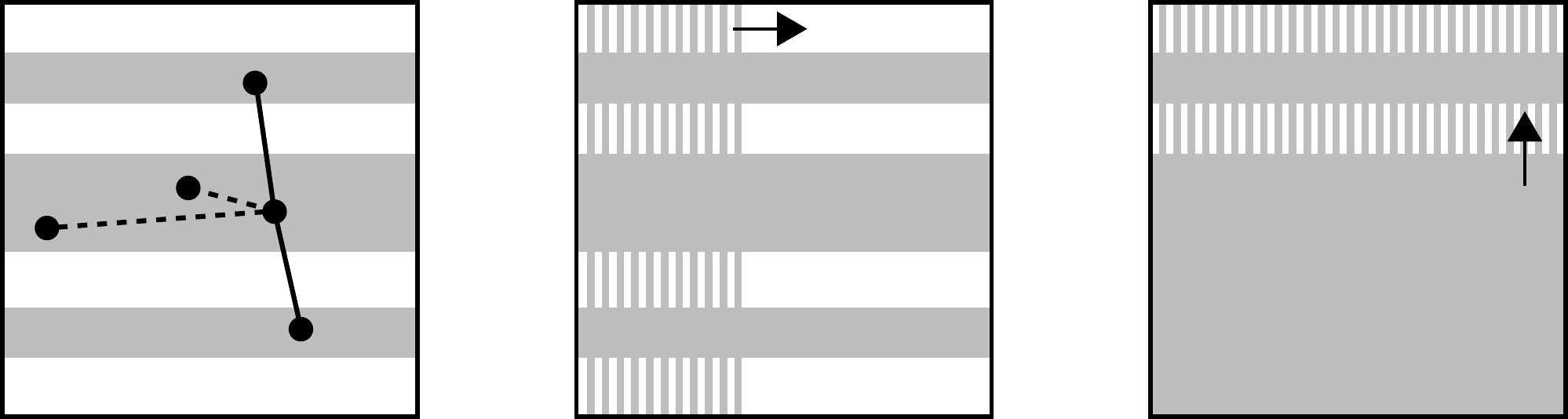}
  \caption{The three stages of embedding the points in $S_\pi$. All points within shaded areas are embedded. In the first stage a point is shown whose neighbors are also embedded. \label{fig3}}
\end{figure*}

In the \emph{first stage} we embed $s$ of the $2s$ sets $P_1, \dots, P_{2s}$, chosen such as to minimize the size of the boundary at the end of the stage. Let $Q^I$ denote the set of points embedded in this stage. In the \emph{second stage} we embed either $Q^E \setminus Q^I$ or $Q^O \setminus Q^I$, depending on which is more advantegous, as explained later. In the \emph{third stage} we embed all remaining points of $S_\pi$ in their increasing order of value. See Figure~\ref{fig3} for an illustration. We now provide more detail and analysis.

\subparagraph*{First stage. } Let $\{i_1, \dots, i_s\} \subset [2s]$ be the collection of $s$ indices for which $Q^I = P_{i_1} \cup \cdots \cup P_{i_{s}}$ has the smallest boundary. Finding these indices in a na\"ive way amounts to verifying all $2s \choose s$ choices, each in linear time. The cost of this step is absorbed in our overall bound on the running time. We then embed $Q^I$ (i.e.\ the first $k/2$ entries of $\tau$ are the indices of points in $Q^I$).

We claim that $|\bd{(Q^I})|$ is at most $k(\frac{1}{2} - \frac{1}{8}) + o(k)$. 
To see this, consider the expected boundary size of $Q^I$ if $\{i_1, \dots, i_s\}$ were a subset of size $s$ of $[2s]$ chosen uniformly at random. A point $p \in Q^I$ is \emph{not} on the boundary, if all neighbors of $p$ are in $Q^I$. Observe that $N^U(p)$ and $N^D(p)$ are in $Q^I$, unless $p.y$ is at the margin of one of the chosen intervals $I_{i_j}$. Thus, at least $\frac{k}{2} - 2s$ points have these two neighbors covered, for all choices of $Q^I$. We now look at the probability that the other two neighbors, $N^L(p)$ and $N^R(p)$ are in $Q^I$. The least favorable case is when the two are in separate intervals, different from the interval of $p$. Fixing all three intervals leaves ${2s-3 \choose s-3}$ choices for completing the selection, out of ${2s-1  \choose s-1}$ choices when only the interval of $p$ is fixed. Thus, the probability that $N^L(p), N^R(p) \in Q^I$, conditioned on $p \in Q^I$ is at least ${2s-3 \choose s-3} / {2s-1  \choose s-1} = \frac{1}{4} - O(\frac{1}{s})$, and therefore the expected number of points in $Q^I$ whose neighbors are all in $Q^I$ is at least $(\frac{k}{2} - 2s)(\frac{1}{4} - O(\frac{1}{s})) = (\frac{k}{8} - O(\frac{k}{s}))$. 

The expected size of $\bd{(Q^I)}$ is thus at least $\frac{k}{2} - \frac{k}{8} + o(k)$. For the optimal choice of $Q^I$ (instead of random), the boundary size clearly cannot exceed this quantity. 

It remains to decide the actual order in which the points of $Q^I$ are embedded, such as to keep the boundary size below its intended target throughout the process. This can be achieved by first embedding those points that have all their neighbors in $Q^I$, with each such point followed by its four neighbors. In this way, for every (at most) five points added, we increase the boundary by one less than the number of points added. This continues until the entire saving of $\frac{k}{8} - o(k)$ is realized. No intermediate boundary size can therefore exceed $\max\{\frac{k}2 - \frac{k}{8}, \frac{4}{5} \cdot \frac{k}{2}\} + o(k) \leq 0.4k + o(k)$.

\subparagraph*{Second stage. }

Let $e^E$, $e^O$ denote the number of points in $Q^E \cap \bd{(Q^I)}$, resp.\ $Q^O \cap \bd{(Q^I)}$, i.e.\ the even- and odd-index points on the boundary of $Q^I$ (call such points \emph{exposed}). Let $h^E$, $h^O$ denote the number of points in $(Q^I \cap Q^E) \setminus \bd{(Q^I)}$, resp.\ $(Q^I \cap Q^O) \setminus \bd{(Q^I)}$, i.e.\ the even- and odd-index points of $Q^I$ \emph{not} on the boundary (call such points \emph{hidden}). 

Observe that $e^E + e^O + h^E + h^O = k/2$, since the sets in question partition $Q^I$. Moreover, $h^O + h^E \geq \frac{k}{8} - o(k)$, by our earlier upper bound for $|\bd{(Q^I)}|$.

If $h^O \geq h^E$, then, in the second stage, we embed all points in $Q^O \setminus Q^I$, i.e.\ all odd-index points not yet embedded. Otherwise, we embed all points in $Q^E \setminus Q^I$, i.e.\ all even-index points not yet embedded. Suppose that we are in the first case. 

At the end of the stage we will have embedded a total number of $\frac{k}{2} + (\frac{k}2 - e^O - h^O) = \frac{k}{2} + e^E + h^E$ points. (The first term counts the points embedded in the first stage, the second term counts the odd-index points \emph{not embedded} in the first stage.

Observe that after this stage, all points in $Q^I \cap Q^E$ are hidden, except for (at most) $2s$ points with values at the margins of intervals $I_{i_j}$. This is because all non-margin points $p \in Q^I \cap Q^E$ have their neighbors $N^U(p),N^D(p)$ already embedded in the first stage, since they fall in the same interval, and neighbors $N^L(p), N^R(p)$ embedded in the second stage, since these are odd-index points.

As there are at least $e^E - 2s$ newly hidden points in the second stage, the \emph{increase} in boundary size is at most $(e^E + h^E) - (e^E - 2s) \leq h^E + 2s$. By the assumption that $h^O \geq h^E$, we have $h^E \leq \frac{k}{16} + o(k)$, the boundary size at the end of the stage is thus at most $|\bd{(Q^I)}| + \frac{k}{16} + o(k) \leq \frac{k}{2} - \frac{k}{16} + o(k) = 0.4375k + o(k)$. We observe that this step is the bottleneck of the entire argument.

Again, we have to show that \emph{during} the second stage, the boundary size never grows above this bound.
To achieve this, we embed first, for each even-index point that is to be hidden in this stage, its two neighbors. As an effect, the entire saving is realized in the beginning of the stage, while the boundary size may grow by at most $h^E + 2s \leq \frac{k}{16} + o(k)$, i.e.\ it stays below the final bound. We can embed the remaining odd-index points in the natural, left-to-right order. In the case when $h^E > h^O$, the procedure and its analysis are symmetric, i.e.\ we embed all points in $Q^E \setminus Q^I$. 

\subparagraph*{Third stage. }

Finally, we embed all remaining points in increasing order of value. (Assuming that $h^O \geq h^E$ was the case in the second stage, this means embedding all points in $Q^E \setminus Q^I$.)

We claim that the boundary does not increase during the process (except possibly by one). To see this, consider the embedding of a point $p \in Q^E \setminus Q^I$. Observe that $N^D(p)$ is already embedded; if it was not embedded in the first two stages, then it must be an even-index point, preceeding $p$ by value, so it must have been embedded in the third stage. Neighbors $N^L(p)$ and $N^R(p)$ have odd index, they were thus embedded in the first two stages. Neighbor $N^U(p)$ either has odd index (is therefore already embedded), or has even index, in which case it will be the next point to be embedded, thereby hiding $p$. This concludes the analysis.

\subsection{Special patterns}\label{sec:spec}

We show that if the pattern $\pi$ is a Jordan-permutation, then PPM can be solved in subexponential time (Theorem~\ref{thm4}). The proof is simple: the incidence graph of Jordan permutations is by definition planar. To see this, recall that a Jordan permutation is defined by the intersection pattern of two curves. We view the curves as the planar embedding of $G_\pi$. The portions of the curves between intersection points correspond to edges (we trim the loose ends of both curves), and the curves connect the points in the order of their index, resp.\ value. We observe that this is in fact, an exact characterization: $G_\pi$ is planar if and only if $\pi$ is a Jordan permutation.\footnote{For the ``only if'' direction, we need to allow \emph{touching points} between the two curves. Consider any noncrossing embedding of $G_\pi$, and construct the two curves as the Hamiltonian paths of $G_\pi$ that connect the vertices by increasing index, resp.\ value. Whenever the two curves overlap over an edge of $G_\pi$, we bend the corresponding part of one of the curves, such as to create two intersection points at the two endpoints of the edge (one of the two intersection points may need to be a touching point).}

The pathwidth of a $k$-vertex planar graph is well-known to be $O(\sqrt{n})$~\cite{DowneyFellows}. A corresponding embedding order $\tau$ of $\pi$ can be built recursively, concatenating the sequences obtained on the different sides of the separator and the sequence obtained from the separator itself. Theorem~\ref{thm4} follows.

It would be interesting to obtain other classes of permutations whose incidence graphs are minor-free. Guillemot and Marx~\cite{GM}  show that a permutation $\pi$ that avoids a pattern of length $\ell$ has a certain decomposition of width $f(\ell)$. To show Conjecture~\ref{conj}, one may relate this width with the size of a forbidden minor in $G_\pi$. We point out that such a connection cannot be \emph{too strong}: we exhibit a $k$-permutation that avoids a fixed pattern, but whose incidence graph contains a large grid, with resulting pathwidth $\Theta(\sqrt{k})$.

\begin{figure*}[h]
  \centering
  \includegraphics[width=11cm]{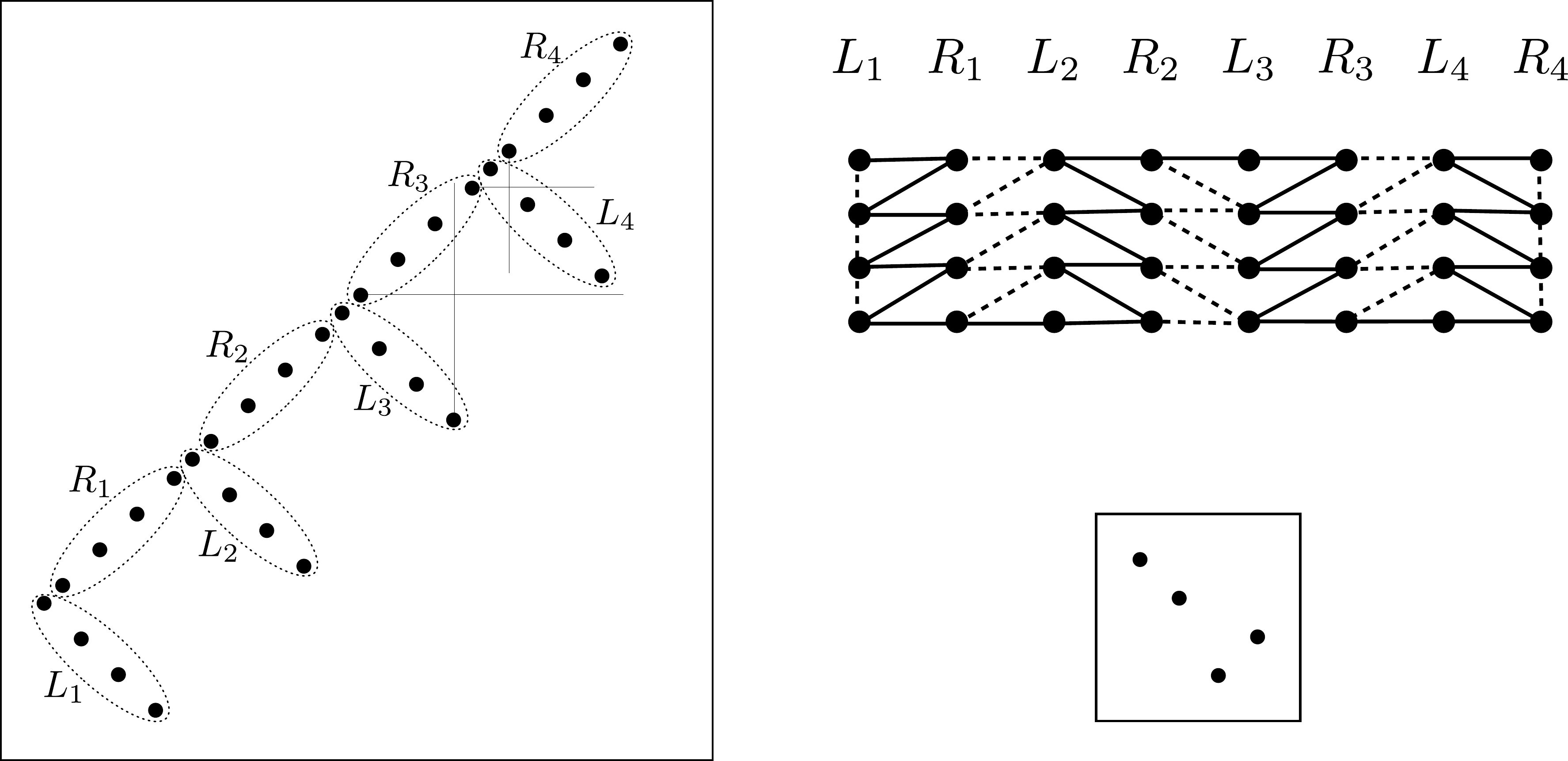}
  \caption{(left) Permutation $\pi(8,4)$, with subsequences generated by $L_i$ and $R_i$. (right, above) Incidence graph of $\pi(8,4)$, with points corresponding to $L_i$, $R_i$ re-arranged in columns, to highlight grid structure. Solid line indicates neighborhood by index, dashed line indicates neighborhood by value. (right, below) Pattern $(4,3,1,2)$ avoided by $\pi(8,4)$. \label{fig5}}
\end{figure*}

Let $a$ and $b$ be parameters such that $a$ is even, and $a b = k$. Let $L_i$ be the sequence of \emph{even} integers in $[1 + (i-1) \cdot a,~i \cdot a]$ in \emph{decreasing} order, and let $R_i$ be the sequence of \emph{odd} integers in $[1 + i \cdot a,~ (i+1) \cdot a]$ in \emph{increasing} order, for $1 \leq i \leq b$. Observe that $|L_i| = |R_i| = a/2$ for all $i$, and that all sequences are disjoint. Let $\pi = \pi(a,b)$ denote the unique permutation of length $k$ that has the same ordering as the concatenation of $T_1,\dots,T_b$, where $T_i$ is obtained from interleaving $L_i$ and $R_i$. (See Figure~\ref{fig5}.)

We observe that $\pi$ avoids the pattern $\sigma = (4,3,1,2)$. To see this, suppose for contradiction that $\pi$ contains $\sigma$. Denote the embeddings of points of $S_\sigma$, by increasing index, as $p_1, p_2, p_3, p_4$. Point $p_2$ must be in one of the $L_i$ sets, for otherwise it could have no point $p_1$ above and to the left. Then, both $p_3$ and $p_4$ must be in the same $L_i$, as only this subset contains points below and to the right of $p_2$. However, no set $L_i$ contains two points in the same relative position as $p_3$, $p_4$ (i.e.\ one above and to the right of the other), a contradiction. Finally, we claim that $G_\pi$ contains a grid of size $\Theta(a) \times \Theta(b)$, as illustrated in Figure~\ref{fig5}.
 
\section{Concluding remarks} \label{sec:open}

It is conceivable that the approaches presented here can be combined with previous techniques, to obtain further improvements in running time. In particular, our bounds depend on $k$ and $n$ only, but one could also consider finer structural parameters. A good understanding of which patterns are easiest to find is still lacking. Conjecture~4 points at a possible step in this direction.

Obtaining tighter bounds for the pathwidth of permutation incidence graphs (and more generally, for the pathwidth of $4$-regular graphs) is an interesting structural question in itself. For $n$-vertex \emph{cubic} graphs, the pathwidth is known to be between $0.082 n$ and $0.167 n$~\cite{FominHoie}. Related expansion-properties are also well-studied, for example, the \emph{bisection-width} of $4$-regular graphs is at most $0.4n + o(n)$~\cite{MonienPreis}. (Bisection-width is a lower bound for pathwidth.)

Finally, as several different approaches for the PPM problem are now known in the literature, a thorough experimental comparison of them would be informative.

\subparagraph*{Acknowledgements.}
An earlier version of the paper contained a mistaken analysis of Algorithm~S. I thank G\"unter Rote for pointing out the error.

This work was prompted by the Dagstuhl Seminar 18451 ``Genomics, Pattern Avoidance, and Statistical Mechanics''. I would like to thank the organizers for the invitation and the participants for interesting discussions.

\newpage

\appendix

\section{Appendix}
\subsection{Proof of Lemma~\ref{eqlem}}
\begin{proof}
Suppose $t$ contains $\pi$, and let $(t(i_1),\dots,t(i_k))$ be the subsequence witnessing this. Let $p_j$ denote the point $(j,\pi(j))$, and set $f(p_j) = (i_j, t(i_j))$ for all $j \in [k]$. Observe that $f(N^L(p_j)).x = i_{j-1}$, and $f(N^R(p_j)).x = i_{j+1}$, the first condition thus holds since $i_{j-1} < i_j < i_{j+1}$. 

Let $\pi(j') = N^D(p_j).y$, and $\pi(j'') = N^U(p_j).y$. By definition, $\pi(j') < \pi(j) < \pi(j'')$. The second condition now becomes $t(i_{j'}) < t(i_j) < t(i_{j''})$, which holds since $t$ contains $\pi$.

In the other direction, let $f:S_\pi \rightarrow S_t$ be a valid embedding.
Define $i_j = f(p_j).x$, for all $j\in[k]$. Since $f(N^L(p_j)).x < f(p_j).x < f(N^R(p_j)).x$ for all $j$, we have $i_1 \leq \cdots \leq i_k$. Let $j',j'' \in [k]$, such that $j' < j''$. 

Then $\pi(i_{j'}) < \pi(i_{j''})$ is equivalent with $t(i_{j'}) = f(p_j).y < f(N^U(\dots(N^U(p_j))\dots)).y = t(i_{j''})$ where the $N^U(\cdot)$ operator, and the second property of a valid embedding are applied $j'' - j'$ times. %The case $j'>j''$ is symmetric.
\end{proof}

%%
%% Bibliography
%%

%% Please use bibtex, 
\newpage

\bibliography{submission}

\appendix

\end{document}